\documentclass[aps,pra,showpacs,amsmath,amssymb,amsfonts,tightenlines,twocolumn,lengthcheck,superscriptaddress]{revtex4-1}
\usepackage{amsmath}
\usepackage{amsfonts}
\usepackage{graphicx}
\usepackage{epsfig}
\usepackage{color}
\usepackage{txfonts}
\usepackage[colorlinks,citecolor=blue]{hyperref}

\begin{document}

\title{Tunable Chiral Bound States in a Dimer Chain of Coupled Resonators}
\author{Jing \surname{Li}}
\affiliation{Synergetic Innovation Center for Quantum Effects and Applications, Key Laboratory for Matter Microstructure and Function of Hunan Province, Key Laboratory of Low-Dimensional Quantum Structures and Quantum Control of Ministry of Education, School of Physics and Electronics, Hunan Normal University, Changsha 410081, China}
\author{Jing \surname{Lu}}
\affiliation{Synergetic Innovation Center for Quantum Effects and Applications, Key Laboratory for Matter Microstructure and Function of Hunan Province, Key Laboratory of Low-Dimensional Quantum Structures and Quantum Control of Ministry of Education, School of Physics and Electronics, Hunan Normal University, Changsha 410081, China}
\author{Z. R. \surname{Gong}}
\thanks{Corresponding author}
\email{gongzr@szu.edu.cn}
\affiliation{College of Physics and Optoelectronic Engineering, Shenzhen University, Shenzhen 518060, P. R. China}
\author{Lan \surname{Zhou} }
\thanks{Corresponding author}
\email{zhoulan@hunnu.edu.cn}
\affiliation{Synergetic Innovation Center for Quantum Effects and Applications, Key Laboratory for Matter Microstructure and Function of Hunan Province, Key Laboratory of Low-Dimensional Quantum Structures and Quantum Control of Ministry of Education, School of Physics and Electronics, Hunan Normal University, Changsha 410081, China}

\begin{abstract}
We study an excitation hopping on a one-dimensional (1D) dimer chain of coupled resonators with
the alternate on-site photon energies, which interacts with a two-level emitter (TLE) by a coupling
point or two adjacent coupling points. In the single-excitation subspace, this system not only
possesses two energy bands with propagating states, but also possesses photonic bound states. The
number of bound states depends on the coupling forms between the TLE and the dimer chain. It is found
that when the TLE is locally coupled to one resonator of the dimer chain, the bound-state has
mirror reflection symmetry. When the TLE is nonlocally coupled to two adjacent resonators, three
bound states with preferred direction arise due to the mirror symmetry breaking. By using chirality
to measure the asymmetry, it is found that the chirality of these bound states can be tuned by changing
the energy differences of single photon in the adjacent resonators, the coupling strengths and the
transition energy of the TLE.
\end{abstract}


\date{\today}
\maketitle
\section{Introduction}
The quantum electrodynamics of light-matter interactions in waveguide systems~\cite{PR718,RMP89,RMP95}
has attracted considerable interest. In a waveguide, the electromagnetic field is confined spatially
in two dimensions and propagates along the remaining one, which is called guided modes. The interference
of spontaneously emitted waves from a quantum emitter (QE) and the incident wave leads to total reflection
of single photons in the one-dimensional (1D) waveguide with linear~\cite{Shen05} or nonlinear~\cite{Zhou08}
dispersion relation. One of the most intriguing is the existence of bound states for photons:
the single-photon bound states in continuum~\cite{zhouPRA78(08),gongPRA78(08),dongPRA79(09),TufPRA87(13)},
where the photon is trapped between the QEs or the mirror and a QE;  the single-photon bound states with energies
slightly outside the continuum~\cite{Zhou09,104(10)023602,Zhou13,lujing14,PalmaPRA89,PRA96(17),AhuPRA98},
where a photon is localized and symmetrical around the resonator coupling to the quantum emitter; multiple-photon
bound states~\cite{Shen07,ZhengPRA10,ShiPRA11,Roy11,Shi16}, which give rise to strong correlations between photons.

Nowadays, widespread attention has been paid on giant atoms. Different from a local interaction between QEs
and light field~\cite{Qiao19,zhouPRA85(12),wangPRA89,HengPRA90,LawPRA96,XuPRA95,WangPRA100,LeiPRA100}, giant atoms
nonlocally couples to light field~\cite{Kockum14,Kockum21,Gustafsson14,Manenti17,Delsing19,Sletten19,Andersson20,Bienfait20}
at multiple points. Such unconventional light-matter interaction occurs when the atomic size is comparable or even
larger than the wavelength of the light. The interference effects between these multiple coupling points leads to
unconventional phenomenon such as frequency dependent relaxation and Lamb shift~\cite{Kockum14}, non-Markovian atomic
dissipation~\cite{Guo17,Guo20,Andersson19} and the decoherence free interatomic interaction~\cite{Kockum18,Carollo20}.
The symmetry bound states outside the band are found numerically for a giant atom coupling to a 1D coupled-resonator
waveguide~\cite{Wang20}. And the asymmetric bound state close to an atomic transition frequency is found in a giant
atom interacting with the photonic mode of an energy band~\cite{NoriPRL126}. In this paper, we study a waveguide quantum
electrodynamics system composed of a two-level emitter (TLE) and 1D coupled resonators arranged in a dimer chain due
to their alternate on-site photon energies. Different from the 1D coupled-resonator waveguide with uniform-hopping rates,
this dimer chain possesses two energy bands, and the TLE non-locally coupled to two adjacent resonators of
the dimer chain. To established the relation of bound states with the mirror symmetry, we present the exact analytical
solution of the bound states in real space for arbitrary atomic transition frequency. The asymmetry of bound states stems
from the nonlocal coupling of the quantum emitter to two resonators with different on-site energies. Although we show how
to judge the symmetry or asymmetry of bound states from the mirror symmetry, the properties of single-photon bound states
include more details, for example, how many bound states in this system; whether they all have the same preferred direction
or not? Is it possible for a bound state to localized at the one-side of the symmetry axe? How to tune the asymmetry of the
bound state?

The paper is organized as follows. In Sec.~\ref{Sec:2}, we propose the model describing the
interaction between a TLE and a chain of coupled resonators, and the equations of the probability
amplitudes are presented in single-excitation subspace. In Sec.~\ref{Sec:3}, we derive the
condition for the single-photon bound state and discuss the asymmetry of all bound states by
introducing the chirality. Finally, a summary has been made in Sec.~\ref{Sec:4}.

\begin{figure}[tbp]
\includegraphics[width=0.47 \textwidth]{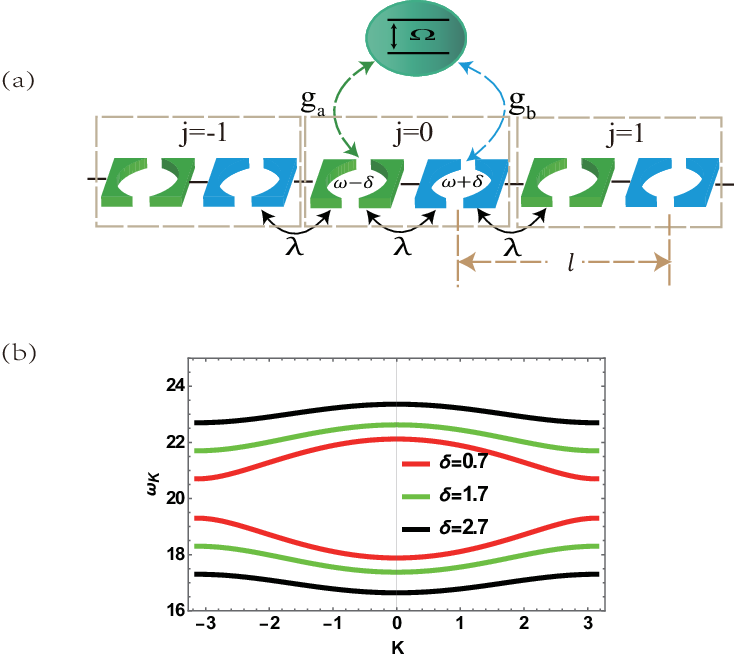}
\caption{(Color online) (a) Schematic depiction of a quantum emitter
nonlocal coupled to the nearest resonators in a 1D dimer chain of coupled
resonators. Dashed boxes indicate the unit cells. (b) The dispersion
relation of the dimer chain for $\protect\delta\neq 0$.}
\label{fig1}
\end{figure}

\section{\label{Sec:2}Model for a quantum emitter nonlocal coupled to a
dimer chain}
We consider a system consisting of a one-dimensional (1D) waveguide and a TLE. The 1D waveguide consists
of a series of coupled resonators in which light propagates due to the coupling between the adjacent
resonators (see Fig.~\ref{fig1}(a)). The system can be implemented by nano-electromechanical resonator
arrays where two nearest resonators with fero-magnetic particles in the tips are coupled to a localized
spin~\cite{Rabl10}, a side-defected cavity with double couplings to a waveguide of coupled defected cavity
arrays~\cite{BabaNP2,Xu10,zhou12}, or a superconducting atoms coupled to a Josephson photonic-crystal
waveguide~\cite{NoriPRL126, Marco22}. In contrast to the previous 1D waveguide with identical resonators~\cite{Zhou08},
the two adjacent resonators (shown as the green and blue cavities in Fig~\ref{fig1}(a)) have on-site photon
energies $\omega_c-\delta$ and $\omega_c+\delta$, respectively. Here, 2$\delta$ denotes the the energy
differences of single photon in the adjacent resonators. We cast this 1D waveguide as a dimer chain of
discrete bosonic sites with equally spaced sites but two kinds of on-site photon energies, and the dimer
chain is assumed to be infinitely long in both direction. Then the continuum of modes in the waveguide are
the Bloch modes. We use $\hat{a}_{2j}$ ($\hat{a}^{\dagger}_{2j}$) and $\hat{b}_{2j+1}$($\hat{b}^{\dagger}_{2j+1}$)
as the bosonic annihilation (creation) operators of a single photon for the green and blue resonators
at the $j$th cell, respectively. Its corresponding real-space Hamiltonian reads
\begin{eqnarray}
\hat{H_c} &=&\left( \omega _{c}-\delta \right) \sum_{j=-\infty}^{+\infty}\hat{a}_{2j}^{\dagger
}\hat{a}_{2j}+\left( \omega _{c}+\delta \right) \sum_{j}\hat{b}%
_{2j+1}^{\dagger }\hat{b}_{2j+1}  \notag \\
&-&\sum_{j}\lambda \left( \hat{a}_{2j}^{\dagger }\hat{b}_{2j+1}+\hat{a}%
^{\dagger }_{2j}\hat{b}_{2j-1}\right)+h.c.  \label{1-01}
\end{eqnarray}
where $\lambda$ is the coupling constant between adjacent resonators in the dimer chain. Two bands of propagating
photons have the following dispersion relation as
\begin{equation}
\omega_{k}^{\pm }=\omega _{c}\pm \sqrt{\delta ^{2}+4\lambda ^{2}\cos
^{2}(k/2)}.  \label{1-02}
\end{equation}
with wave number $k\in[-\pi,\pi]$. The bands of propagating photons with different $\delta$ in the first Brillouin
zone is depicted in Fig.~\ref{fig1}(b). The time-reversal symmetry is satisfied for the propagating modes in the
dimer chain since $\omega^{\pm}_{-k}=\omega^{\pm}_{k}$.

The TLE's ground and excited states $|g\rangle $ and $|e\rangle $, respectively, are separated in energy by $\Omega $.
The transition $|g\rangle \leftrightarrow |e\rangle $ of the TLE is dipole coupled with coupling strength $g_{a}$
($g_{b}$) to the resonator at the $0$th cell. Defining the rising operators $\hat\sigma _{+}=|e\rangle \langle g|$,
and its adjoint $\hat\sigma _{-}$, the Hamiltonian for the free TLE part and the interaction between the TLE and
the field within the rotating-wave approximation reads
\begin{equation}
\hat{H_{1}}=\Omega \left\vert e\right\rangle \left\langle e\right\vert
+\hat\sigma _{+}\left( g_{a}\hat{a}_{0}+g_{b}\hat{b}_{1}\right) +h.c.  \label{1-03}
\end{equation}%
Here, we have introduced the nonlocal interaction between the TLE and the light field.

The number operator $\hat{N}=\sum_{j}(\hat{a}_{2j}^{\dagger }\hat{a}_{j}+\hat{b}_{2j+1}^{\dagger }\hat{b}%
_{2j+1})+|e\rangle \langle e|$ commutes with the total Hamiltonian $\hat{H}=\hat{H}_{c}+\hat{H}_{1}$. We
restrict the analysis to the subspaces with single excitation hereafter. In the single-excitation subspace,
there are two mutual exclusive possibilities: the particle either is propagating inside the cavity or is
absorbed by the TLE. Letting $\left\vert \emptyset\right\rangle =\left\vert 0g\right\rangle $ be the state
without photon while the TLE stays on its ground state, the eigenstate of the Hamiltonian reads%
\begin{equation}
\left\vert \epsilon \right\rangle =\left( \sum_{j}\alpha _{2j}\hat{a}%
_{2j}^{\dagger }+\sum_{j}\beta _{2j+1}\hat{b}_{2j+1}^{\dagger }+u_{e}\hat{%
\sigma}_{+}\right) \left\vert \emptyset \right\rangle ,  \label{1-04}
\end{equation}%
where $\alpha _{2j}$, $\beta _{2j+1}$ are the probability amplitudes to find a photon in $a$ and $b$ resonators
of the jth cell, respectively, and $u_{e}$ is the probability amplitude of the TLE in the excited state while
no photon in the dimer chain. From the stationary Schr\"{o}dinger equation, one can obtain the equations for the
amplitudes. By removing $u_{e}$, the equations for the photonic amplitudes reduce to
\begin{subequations}
\label{1-06}
\begin{eqnarray}
&&\left( \epsilon -\omega _{c}+\delta \right) \alpha _{2j} \\
&=&-\lambda \left( \beta _{2j+1}+\beta _{2j-1}\right) +\delta _{j0}\left(
V_{a}\alpha _{0}+G\beta _{1}\right) ,  \notag \\
&&\left( \epsilon -\omega _{c}-\delta \right) \beta _{2j+1} \\
&=&-\lambda \left( \alpha _{2j}+\alpha _{2j+2}\right) +\delta _{j0}\left(
G^{\ast }\alpha _{0}+V_{b}\beta _{1}\right) ,  \notag
\end{eqnarray}%
which lead to a nonlocal energy-dependent delta-like potentials $V_{n}$, $%
n=a,b$ and the effective dispersive coupling strength
\end{subequations}
\begin{equation}
V_{n}=\frac{g_{n}^{\ast }g_{n}}{\epsilon -\Omega },G=\frac{g_{a}^{\ast }g_{b}%
}{\epsilon -\Omega }.  \label{1-07}
\end{equation}%
Obviously, the effective dispersive coupling strength $G$ vanishes when the TLE only
interacts with one resonator of the unit cell, which plays an important role in the emergence of the chiral bound states.

\begin{figure*}[tbp]
\includegraphics[width=1 \textwidth]{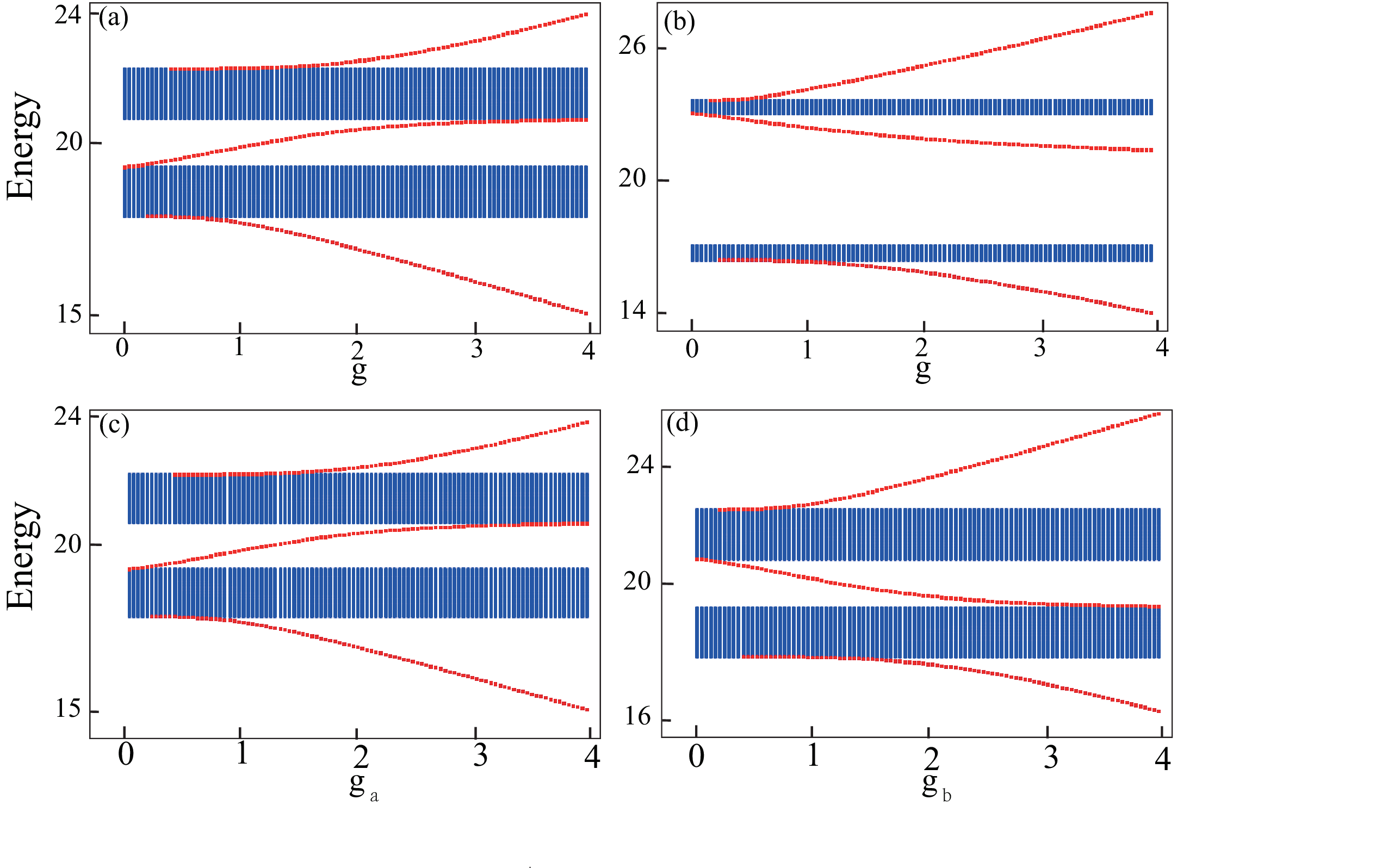}
\caption{(Color online) The energy versus coupling strength $g_{a}=g_{b}=g$
in (a) and (b), $g_{a}$ in (c) and $g_{b}$ in (d) for $\protect\omega _{c}=20
$. The other parameters are setting as follow: (a) $\protect\delta %
=0.7,\Omega =19.3$, (b) $\protect\delta =3,\Omega =23$, (c) $g_{b}=0,\protect%
\delta =0.7,\Omega =19.3$, (d) $g_{a}=0,\protect\delta =0.7,\Omega =20.7$.
All parameters are in units of the hop strength $\protect\lambda $.}
\label{fig2}
\end{figure*}

\section{\label{Sec:3} Single-photon Bound States}

The presence of the TLE breaks down the translational symmetry of the dimer chain, which leads to the
highly-localized states. We plot the energy spectrum versus coupling strengths in the single excitation
subspace in Fig.~\ref{fig2} by numerical diagonalization of the Hamiltonian in the real space with 100 resonators. The periodic boundary conditions is used. Two energy
bands of scattering states are symmetrically formed above and below $\epsilon =\omega _{c}$ with $2\delta $
as the band gap. Obviously, the two bands merge to one band when there is no energy difference between
resonators ($\delta=0$). The three curves, one above all bands, one below all bands and the other inside
the band gap, are the bound states. As coupling strengths increase, the energy differences between the bound
states and the band edges also increase. However, the increment of the energy differences between the bound
states and the band edges is dependent on transition frequency $\Omega$, especially for the emerging
bound state inside the gap. When $\Omega <\omega _{c}$, the energy of the emerging bound state increases
as the coupling strengths increase. When $\Omega >\omega _{c}$, the energy of the emerging bound state
decreases as the coupling strengths increase. In Fig.~\ref{fig2}(c) and (d) the TLE only interacts with
one resonator of the unit cell. It is noted that Fig.\ref{fig2}(d) can be obtained by the mirror reflection
of the Fig.\ref{fig2}(c) with mirror surface locating at $\epsilon =\omega _{c}$, which actually indicates
the same mirror reflection symmetry of the dimer resonator energies and the TLE's energy when $a \rightarrow b $
and $ \Omega-\omega_{c} \rightarrow -(\Omega-\omega_{c})$. Hereafter, for brevity, the bound state above,
below the bands are denoted bound state I, II, and the bound state in the band gap is denoted as bound state III.

The highly-localized states in previous studies~\cite{Zhou09,104(10)023602,Zhou13,lujing14,PalmaPRA89,PRA96(17),AhuPRA98,Wang20}
decay exponentially and symmetrically in both directions, actually, the symmetry of the bound state is guaranteed
by the mirror reflection symmetry around the quantum emitter. However, the mirror reflection symmetry is broken in
our model with all $g_{n}\neq 0$. Since only the energy bands are modified, there should still be localized modes
of photons around the cell where the TLE is embedded. So we assume the following damped wave
\begin{subequations}
\label{2-01}
\begin{eqnarray}
\alpha _{2j} &=&\left\{
\begin{array}{c}
A_{L}^{\kappa }e^{ik_{0}j+\kappa j},j<0 \\
A_{R}^{\kappa }e^{ik_{0}j-\kappa j},j>0%
\end{array}%
\right.   \\
\beta _{2j+1} &=&\left\{
\begin{array}{c}
B_{L}^{\kappa }e^{ik_{0}j+\kappa \left( j+1/2\right) },j<0 \\
B_{R}^{\kappa }e^{ik_{0}j-\kappa \left( j+1/2\right) },j>0%
\end{array}%
\right.
\end{eqnarray}%
\end{subequations}
with $k_{0}=0,\pi $, which decreases exponentially with the distance from
the $0$th cell. The imaginary wave vector $\kappa >0$ labels the energy%
\begin{equation}
\epsilon =\omega _{c}\pm \sqrt{\delta ^{2}+\lambda ^{2}\left(
2+e^{ik_{0}+{\kappa} }+e^{-ik_{0}-{\kappa} }\right) }  \label{2-02}
\end{equation}%
of a localized photon outside of the bands. At the region far away from the
TLE, we obtain%
\begin{equation}
\frac{A_{L}^{\kappa }}{B_{L}^{\kappa }}=-\lambda \frac{e^{\kappa
/2}+e^{-ik_{0}-\kappa /2}}{\epsilon -\omega _{c}+\delta },\frac{%
A_{R}^{\kappa }}{B_{R}^{\kappa }}=-\lambda \frac{e^{-ik_{0}+\kappa
/2}+e^{-\kappa /2}}{\epsilon -\omega _{c}+\delta },  \label{2-03}
\end{equation}%
By applying Eqs.(\ref{2-01}) and (\ref{2-03}) to the discrete scattering
equation at $j=\pm 1$, the amplitudes for the $j=0$ cell can be obtained as%
\begin{equation}
\alpha _{0}=A_{L}^{\kappa },\beta _{1}=B_{R}^{\kappa }e^{-\kappa /2}.
\label{2-04}
\end{equation}%
Substituting Eq.(\ref{2-04}) and the spatial exponential-decay solution (\ref%
{2-01}) to Eq. (\ref{1-06}) at $j=0$ yields the condition for the energy of the bound state
\begin{eqnarray}
\left\vert G-\lambda \right\vert ^{2} &=&\left( \epsilon -\omega _{c}+\delta
-V_{a}+\lambda \frac{B_{L}^{\kappa }}{A_{L}^{\kappa }}e^{-ik_{0}-\kappa
/2}\right)  \label{2-06} \notag\\
&&\times \left( \epsilon -\omega _{c}-\delta -V_{b}+\lambda \frac{%
A_{R}^{\kappa }}{B_{R}^{\kappa }}e^{ik_{0}-\kappa /2}\right)
\end{eqnarray}%
Once $\kappa $ is obtained from Eq.(\ref{2-06}), so does the ratio in Eq.(%
\ref{2-03}). To give an intuitional knowledge on the bound state, we rewrite
the wave function of the bound states as
\begin{subequations}
\label{2-07}
\begin{align}
\frac{\alpha _{2j}}{A_{R}^{\kappa }}& =\left\{
\begin{array}{c}
\frac{A_{L}^{\kappa }}{A_{R}^{\kappa }}e^{ik_{0}j+\kappa j},j<0 \\
\frac{A_{L}^{\kappa }}{A_{R}^{\kappa }},\text{ }j=0 \\
e^{ik_{0}j-\kappa j},\text{ }j>0%
\end{array}%
\right. \\
\frac{\beta _{2j+1}}{A_{R}^{\kappa }}& =\left\{
\begin{array}{c}
\frac{A_{L}^{\kappa }}{A_{R}^{\kappa }}\frac{B_{L}^{\kappa }}{A_{L}^{\kappa }%
}e^{ik_{0}j+\kappa \left( j+1/2\right) },j<0 \\
\frac{B_{R}^{\kappa }}{A_{R}^{\kappa }}e^{-\kappa /2},\text{ }j=0 \\
\frac{B_{R}^{\kappa }}{A_{R}^{\kappa }}e^{ik_{0}j-\kappa \left( j+1/2\right)
},\text{ }j>0%
\end{array}%
\right.
\end{align}%
where the ratio
\end{subequations}
\begin{equation}
\frac{A_{L}^{\kappa }}{A_{R}^{\kappa }}=\frac{\lambda }{\lambda -G^{\ast }}-%
\frac{V_{b}e^{-\kappa /2}}{\left( G^{\ast }-\lambda \right) }\frac{%
B_{R}^{\kappa }}{A_{R}^{\kappa }}.  \label{2-08}
\end{equation}

When $k_{0}=0$, Eq.(\ref{2-06}) gives the energy $\epsilon _{o}^{\pm
}=\omega _{c}\pm \sqrt{\delta ^{2}+4\lambda ^{2}\cosh ^{2}\left( \kappa
/2\right) }$ lying upper or below all the bands, respectively. When $%
k_{0}=\pi $, Eq.(\ref{2-06}) shows that the energy $\epsilon _{m}^{\pm
}=\omega _{c}\pm \sqrt{\delta ^{2}-4\lambda ^{2}\sinh ^{2}\left( \kappa
/2\right) }$ is inside the band gap. The ratios in Eq.(\ref{2-03}) indicate
that
\begin{subequations}
\label{2-09}
\begin{eqnarray}
\frac{A_{L}^{\kappa }}{B_{L}^{\kappa }} &=&\frac{A_{R}^{\kappa }}{%
B_{R}^{\kappa }}\text{ for }k_{0}=0 \\
\frac{A_{L}^{\kappa }}{B_{L}^{\kappa }} &=&-\frac{A_{R}^{\kappa }}{%
B_{R}^{\kappa }}\text{ for }k_{0}=\pi
\end{eqnarray}%
\end{subequations}
When the TLE only couples to the resonator at the $0$th site, $A_{L}^{\kappa
}/A_{R}^{\kappa }=e^{-ik_{0}}$. When TLE only couples to the resonator at the $1$th
site, $A_{L}^{\kappa }/A_{R}^{\kappa }=e^{-ik_{0}+\kappa }$. In Fig.~\ref%
{fig3}, we have plotted the photonic probability distribution of the bound
state I (a,d,g), III (b,e,h) and II (c,f,i).
\begin{figure*}[tbp]
\includegraphics[width=1 \textwidth]{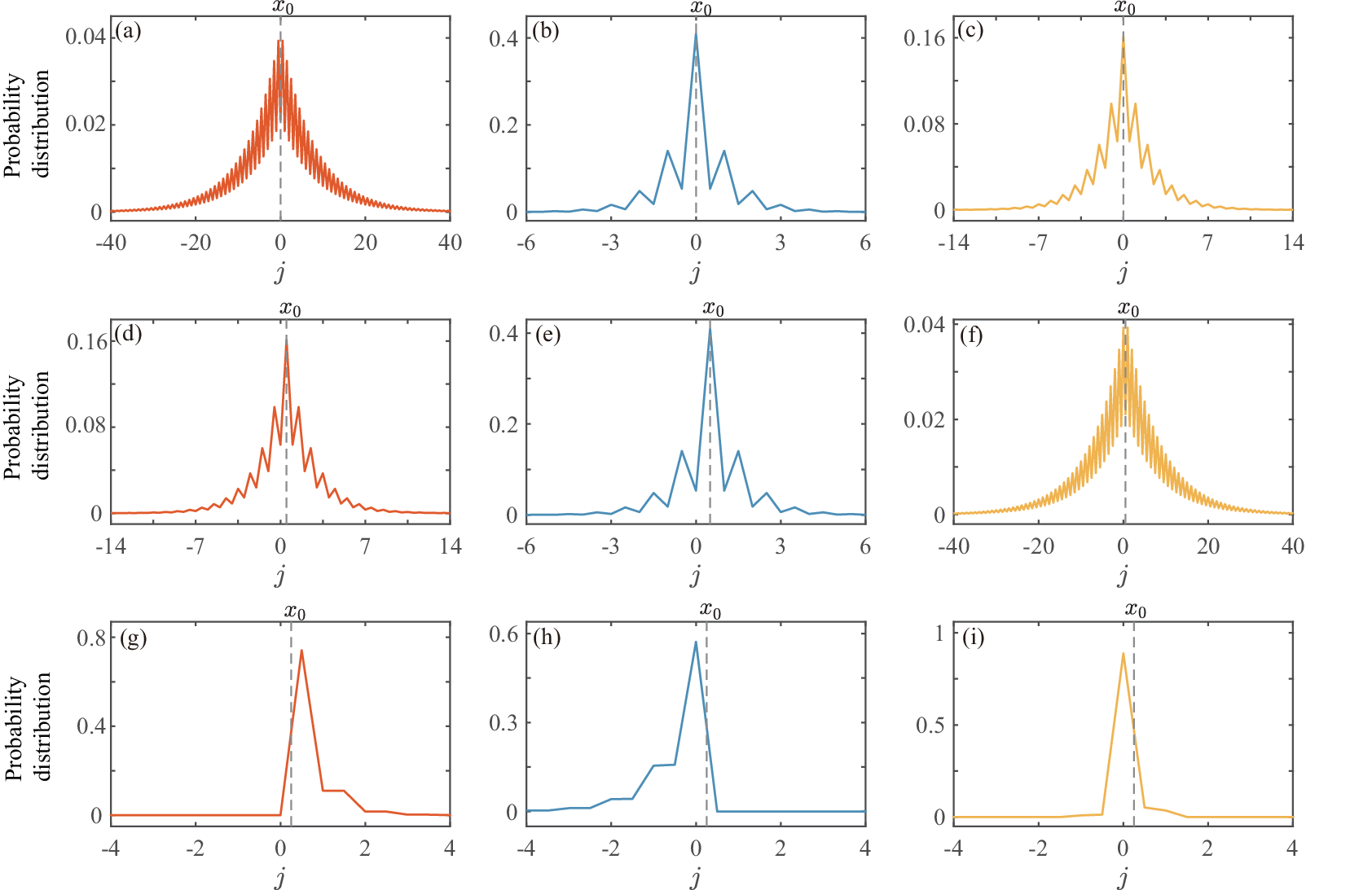}
\caption{(Color online) The bound state distribution above all the bands
(a,d,g), inside the band gap (b,e,h) and below all the bands (c,f,i). In
panels (a-c), $g_{a}=0.5,g_{b}=0,\delta=0.7,\Omega =19.3$, i.e., the TLE only interacts
with the resonator at the $0$th site. In panels (d-f), $g_{a}=0,g_{b}=0.5,%
\delta=0.7,\Omega =20.7$, the TLE only interacts with the resonator at the $1$th site.
In panels (g-i), the TLE interacts with the adjacent resonators at the $0$th
cell, $g_{a}=g_{b}=0.7$, (g) $\Omega=22, \delta=1.1$ (h) $\Omega=19.3,\delta=0.7$,
(i) $\Omega=15.5, \delta=3.8$. All parameters are in units of the hop strength
$\lambda$ and $\omega _{c}=20$.}
\label{fig3}
\end{figure*}
It can be found from Fig.~\ref{fig3}(a-f) that all the bound states are
symmetry about the symmetry axis $x_0$. For different coupling situations, the position of the symmetry axis $x_0$ is different. When the TLE only interacts with single resonator,
the same mirror reflection symmetry of the dimer resonator energies and the TLE's energy under $a \rightarrow b $ and $ \Omega-\omega_{c} \rightarrow -(\Omega-\omega_{c})$
still can be found  in panels (a-c) and (d-f) (e.g. Fig.~\ref{fig3}(a) and Fig.~\ref{fig3}(f)). However, when the TLE interacts with two adjacent resonators
at one unit cell, the bound states no longer has the mirror reflection symmetry as shown in Fig.~\ref{fig3}%
(g-i).

If the TLE is located at the middle of the unit cell at $j=0$, e.g. position $
x_{0}=l/4$, the line at the TLE's location divides the space into left- and
right-hand side of the TLE. The photonic component of the bound state in Fig.~\ref%
{fig3}(g) are strongly localized at the right-hand side of the TLE, and that in Fig.~\ref{fig3}(h)
mostly distributes to the left-hand side of the TLE, the asymmetry of the photonic
probability on both side of the $x_0$ axis can also be found in Fig.~\ref{fig3}(i).
According to the reflection symmetry in geometry, we introduce the chirality~\cite{NoriPRL126}
\begin{equation}
S=\frac{s_{L}-s_{R}}{s_{L}+s_{R}} \label{2-10}
\end{equation}%
to depict the asymmetry of the bound state in left- and right-hand side of the TLE
\begin{subequations}
\label{2-11}
\begin{eqnarray}
s_{L} &=&\sum_{j=-1}^{-\infty }\left( \left\vert \alpha _{2j}\right\vert
^{2}+\left\vert \beta _{2j+1}\right\vert ^{2}\right) +\left\vert \alpha
_{0}\right\vert ^{2}, \\
s_{R} &=&\sum_{j=1}^{\infty }\left( \left\vert \alpha _{2j}\right\vert
^{2}+\left\vert \beta _{2j+1}\right\vert ^{2}\right) +\left\vert \beta
_{1}\right\vert ^{2},
\end{eqnarray}%
\end{subequations}
where $S>0$ ($S<0$) indicates that the left-handed (right-handed) chirality, and $S\rightarrow 1$ ($S\rightarrow -1$)
corresponds to the perfect left-handed (right-handed) chirality. By applying the wave
function in Eq.(\ref{2-07}), we can obtain the expression of the charity in terms of
the ratios%
\begin{equation}
S=\frac{\left\vert A_{L}^{\kappa }/A_{R}^{\kappa }\right\vert ^{2}+\left(
\left\vert A_{L}^{\kappa }/A_{R}^{\kappa }\right\vert ^{2}-1\right) \frac{%
e^{-\kappa }+\left\vert B_{L}^{\kappa }/A_{L}^{\kappa }\right\vert ^{2}}{%
1-e^{-2\kappa }}e^{-\kappa }}{\left\vert A_{L}^{\kappa }/A_{R}^{\kappa
}\right\vert ^{2}+\left( \left\vert A_{L}^{\kappa }/A_{R}^{\kappa
}\right\vert ^{2}+1\right) \frac{e^{-\kappa }+\left\vert B_{L}^{\kappa
}/A_{L}^{\kappa }\right\vert ^{2}}{1-e^{-2\kappa }}e^{-\kappa }} \label{2-12}
\end{equation}

\begin{figure*}[tbp]
\includegraphics[width=1\textwidth]{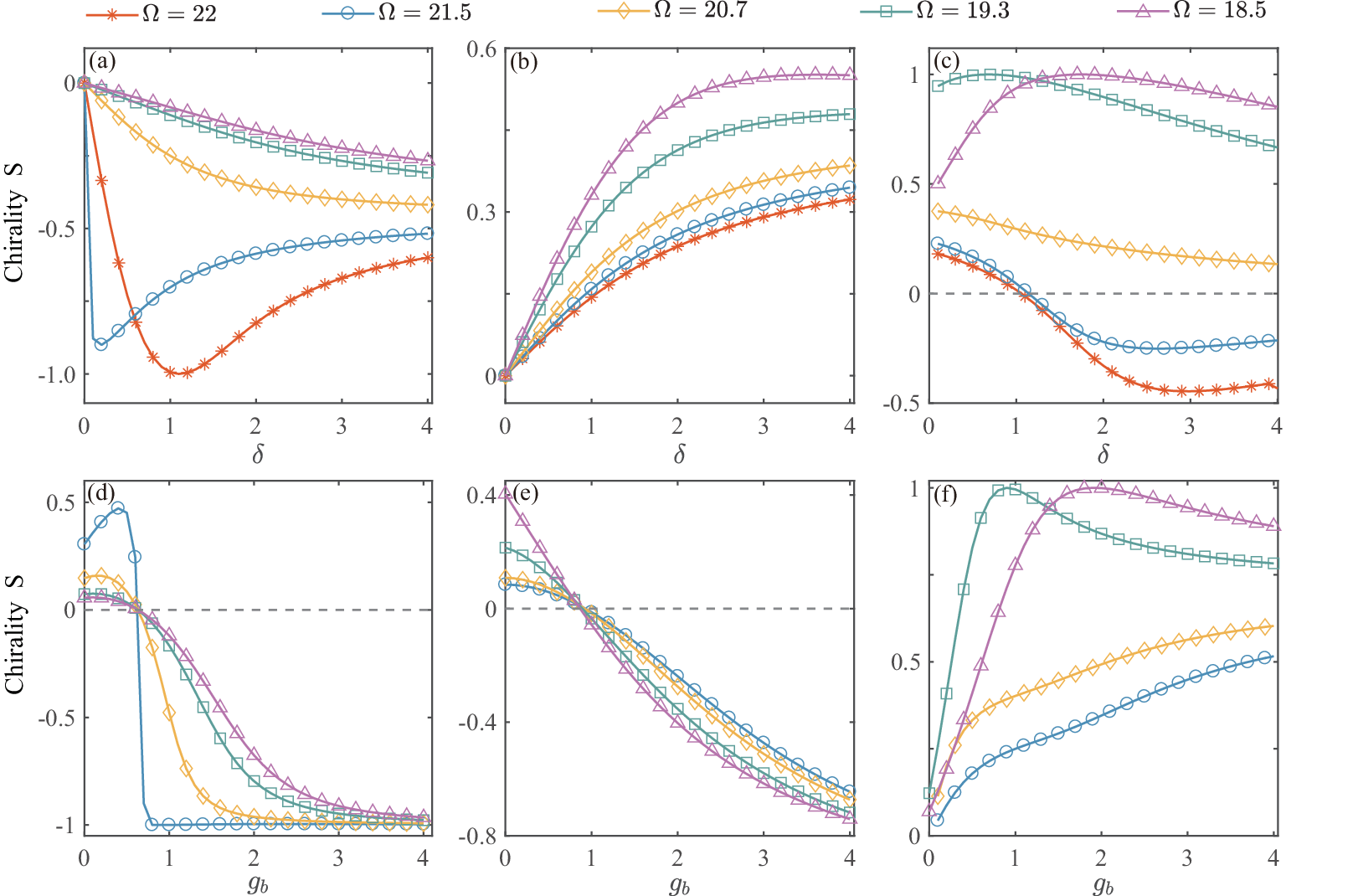}
\caption{(Color online) Tuning the chirality of the bound state above all
the bands (a,d), below all the bands (b,e) and inside the band gap (c,f).
(a)-(c) Chirality versus $\protect\delta $ for fixed coupling strength
$g_{a}=g_{b}=0.7$, and $\Omega=22 $(red),$\Omega =21.5$ (blue), $\Omega =20.7$ (yellow),
$\Omega =19.3$ (green) and $\Omega =18.5$ (purple). (d)-(f) Chirality
versus $g_{b}$ for fixed $\delta =0.2$, $g_{a}=0.7$, $\Omega =21.5$ (blue),
$\Omega =20.7$ (yellow), $\Omega =19.3$ (green) and $\Omega =18.5$ (purple). All parameters are in units of the hop strength $\lambda$ and $\omega _{c}=20$.}
\label{fig4}
\end{figure*}
To show the dependence of the chirality on the parameters, we have plotted the
chirality $S$ of the bound state versus $\delta $ in Fig.~\ref{fig4}(a-c)
and the coupling strength in Fig.~\ref{fig4}(d-f). Since the TLE-resonator
coupling strengths are equal in panels (a-c), the break of the reflection
symmetry depends on whether $\delta $ vanishes or not. It is shown that all chirality
vanishes at $\delta =0$ in panels (a,b) for ordinary bound states I and II. While since
the merging bound state III disappear for $\delta =0$, the critical chirality of the
emerging bound state III in panel (c) is quite different for different transition energies of
the TLE, while the chirality of ordinary bound states tends to coincidence. One can tune the chirality by adjusting the transition energy $%
\Omega $ for a given $\delta $: the left-handed chirality decreases as $\Omega $
increases for bound state II, the right-handed chirality can increase as $\Omega $
increases under certain parameters for bound state I. It should be aware that the chirality of bound
state I prefers the right-hand side of dimer chain and bound state II prefers the left
hand side of dimer chain. Although the chirality of all bound states can be continuously
tuned by adjusting $\delta $, the right
(left) hand side chirality remains unchanged for bound state I (II), only bound state
III can change its sign of chirality for an appropriate $\Omega $ (see the
green line in panel c). The perfect right (left) chirality can be achieved
only for the bound state I (III). For a given $\delta $, one can change the
sign of chirality of the bound states I and II by adjusting one of the
coupling strengths, see Fig.~\ref{fig4}(d) and (e). These changes can be
easily explained by following reasons: When $g_{b}\ll g_{a}$, bound states are
mainly localized around the $0$th site, however the symmetry line is
positioned at $x_{0}=l/4$ not $x_{0}=0$, they present left handed chirality; When $%
g_{b}\gg g_{a}$, bound states are mainly localized around the $1$th site,
thence the symmetry line positioned at $x_{0}=l/4$ gives rise to a right handed
chirality of bound state I and II. It can be also found that the left handed chirality
increases (decreases) as $\Omega $ increases when $g_{b}$ is smaller than $%
g_{a}$ and the right handed chirality increases (decreases) as $\Omega $ increases
when $g_{b}$ is larger than $g_{a}$ for bound state I (II), the perfect
right (left) charity can be still achieved only for the bound state I (III).

\section{\label{Sec:4} Conclusion}

In summary, we consider the TLE is coupled to two adjacent resonators, where two resonators
in each dimer have different on-site photon energies. In the single-excitation
subspace, this system has propagating states forming two energy bands and as well as bound-states.
We mainly focus on the chiral
feature of the bound states. After obtaining the analytical solutions of the bound states in real space, we found that
1) the bound-state distributes symmetrically around the coupling point when the TLE is
locally coupled to one resonator of the dimer chain; 2) the mirror-reflection symmetry breaking
leads to the formation of three chiral bound states when the TLE is nonlocal coupled to two
adjacent resonators. The chirality of the bound states inherits from the geometry aspects,
which are characterized by either the difference of on-site energies in each unit cell or the coupling strengths
between the TLE and the resonators, but it can also be tuned by the transition energy of
the TLE. When the coupling strengths are identical, the nonvanishing difference of on-site
energies lead to right handed chirality of bound state I and left handed chirality of bound state II,
and bound state III can change its preferred chirality by adjusting on-site
difference together with appropriate transition energy of the TLE. For given on-site energies,
one can change the preferred chirality the bound states I and II by adjusting
one of the coupling strengths.

\begin{acknowledgments}
We are grateful to Z.H. Wang for useful discussions.

This work was supported by NSFC Grants No. 11975095, No. 12075082, No.11935006, No. 12175150,
the science and technology innovation Program of Hunan Province (Grant No. 2020RC4047)and the
Natural Science Foundation of Guang-dong Province (Grant No. 2019A1515011400).
\end{acknowledgments}

\end{document}